\documentstyle[12pt]{article}

\newcommand{\sect}[1]{\setcounter{equation}{0}\section{#1}}

\newcommand{\EQ}{\begin{equation}}
\newcommand{\EN}{\end{equation}}
\newcommand{\bea}{\begin{eqnarray}}
\newcommand{\ena}{\end{eqnarray}}

\renewcommand{\a}{\alpha}
\renewcommand{\b}{\beta}
\renewcommand{\c}{\gamma}
\renewcommand{\d}{\delta}
\newcommand{\e}{\epsilon}
\newcommand{\la}{\lambda}

\newcommand{\shalf}{\frac{1}{2}}
\newcommand{\pa}{\partial}
\begin{document}

\topmargin 0pt
\oddsidemargin 5mm

\renewcommand{\Im}{{\rm Im}\,}
\newcommand{\NP}[1]{Nucl.\ Phys.\ {\bf #1}}
\newcommand{\AP}[1]{Ann.\ Phys.\ {\bf #1}}
\newcommand{\PL}[1]{Phys.\ Lett.\ {\bf #1}}
\newcommand{\NC}[1]{Nuovo Cimento {\bf #1}}
\newcommand{\CMP}[1]{Comm.\ Math.\ Phys.\ {\bf #1}}
\newcommand{\PR}[1]{Phys.\ Rev.\ {\bf #1}}
\newcommand{\PRL}[1]{Phys.\ Rev.\ Lett.\ {\bf #1}}
\newcommand{\PTP}[1]{Prog.\ Theor.\ Phys.\ {\bf #1}}
\newcommand{\PTPS}[1]{Prog.\ Theor.\ Phys.\ Suppl.\ {\bf #1}}
\newcommand{\MPL}[1]{Mod.\ Phys.\ Lett.\ {\bf #1}}
\newcommand{\IJMP}[1]{Int.\ Jour.\ Mod.\ Phys.\ {\bf #1}}
\newcommand{\JP}[1]{Jour.\ Phys.\ {\bf #1}}
\renewcommand{\thefootnote}{\fnsymbol{footnote}}

\begin{titlepage}
\setcounter{page}{0}
\rightline{OS-GE 26-92}

\vspace{2cm}
\begin{center}
{\Large Discrete States in Two-Dimensional Quantum Gravity\footnote{An
expanded version of the talk given at the workshop on {\em Elementary
Particles and Symmetries}, Hiroshima University, April 21--23, 1992.}}
\vspace{1cm}

{\large Nobuyoshi Ohta} \\
\vspace{1cm}
{\em Institute of Physics, College of General Education,
Osaka University  \\  Toyonaka, Osaka 560, Japan} \\
\end{center}

\vspace{5mm}
\centerline{{\bf{Abstract}}}

We review the recent developments in the two-dimensional (super)gravity
coupled to $c (\hat c) \leq 1$ (super)conformal matter in
the conformal gauge. Starting from a pedagogical account of the
conformal anomaly in such a system, we show how the system is transformed
into the representation in terms of the (free) Liouville field. Some
perturbative justification is given to this procedure. The physical
states are then examined both for the bosonic and supersymmetric
theories, using the BRST formulation. It is explained how new discrete
states arise together with some examples. We also discuss the relation
with the results for $c = -2$ ``topological" gravity.
The vertex operator representations for the discrete states are
summarized for $c=1$ theory and are used to examine the interactions
of these states. It is found that the states with nontrivial ghost
number have interactions governed by the area-preserving diffeomorphism
similar to those with vanishing ghost number. The resulting effective
action has a BRST-like symmetry.

\end{titlepage}
\newpage
\renewcommand{\thefootnote}{\arabic{footnote}}
\setcounter{footnote}{0}

\sect{Introduction}
\indent

The last few years have witnessed remarkable progress in the attempts
to treat two-dimensional ($2D$) quantum gravity nonperturbatively. This
has been initiated by the discovery of the double scaling limit in the
matrix models, which enables us to go beyond genus expansion by means
of the differential equations satisfied by the nonperturbative
partition function [1-3].

These advances have spurred much progress in the continuum approach
using the Liouville theory. As is well known [4-6], $2D$ gravity coupled
to conformal matter in the conformal gauge reduces to the Liouville
theory with complicated nonlinear dynamics through conformal anomaly.
However, inspired by the recent exact solution of the Liouville
system in the light cone gauge~\cite{POL2}, a method based on
conformal field theory has been well developed and this allows
us to treat the Liouville field as almost free field [8-10].
Various quantities such as correlation functions have been successfully
computed in this approach~\cite{GLI}. The consistency of whole this
procedure requires that the conformal anomaly for the total system
(conformal matter coupled to the Liouville theory) vanish!!
So far this approach makes sense only when the matter conformal field
theory (CFT) has the central charge $c^M\leq 1$.

In this approach, the system may be regarded effectively as a critical
string theory in two dimensions, since the Liouville field provides
a ``time-like" dimension in addition to the space coordinate
representing the conformal matter. It is then expected that there
will be no degrees of freedom beyond that corresponding to the center
of mass since there are no transverse directions; the center
of mass motion of the string gives rise to a scalar particle, which in
our case becomes massless but is usually called ``tachyon" in analogy
to the usual critical string. However, it has been found both in the
matrix model~\cite{GKN} and Liouville approaches [12-19]
that there exist an infinite number of extra degrees of freedom
at discrete values of momenta. It is difficult, if not impossible,
to understand the origin and the role of these ``extra states" in the
matrix models since it is not clear what characterize these states. It
is thus important to try to understand these issues in the continuum
approach. In this article, we will try to give a detailed and
pedagogical account of the Liouville approach to the $2D$ quantum
gravity and, in particular, clarify the origin and interactions of
these states.

In sect.~2, we begin by reviewing how conformal anomaly arises in the
$2D$ gravity coupled to conformal matter. This discussion shows that
the $2D$ gravity
which appears to have no degrees of freedom actually leaves its
trace as the Liouville theory. In sect.~3, we introduce the DDK
ansatz~\cite{DDK} to reduce the theory to the ``free" Liouville theory
coupled to the conformal matter such that the total central
charge vanishes, giving a conformally invariant system. In sect.~4, some
perturbative justification of this trick is discussed~\cite{DHO}.
In sect.~5, we start analyzing the physical states using the BRST
formalism [21-26]. After some preparation in sect.~5.1, we examine
some simple examples in the bosonic theory and give an idea on the
general mechanism of the origin of the extra discrete states in
sect.~5.2. We then discuss the general case in sect.~5.3.
The supersymmetric case is next briefly summarized in sect.~6. Since
the general idea is explained in the bosonic case, we
only sketch the main steps. In sect.~7, we discuss the relation of our
approach to the $c=-2$ theory as considered ``topological gravity"
[27-30]. In sect.~8, we give a summary of the vertex operator
representations of the extra state for $c({\hat c})=1$ theory and
check their BRST invariance. We use these representations in sect.~9
to examine the interactions of these discrete states with {\em and}
without ghost number and show that their interactions are governed by
the symmetry of area-preserving diffeomorphism [15,16,19,31-35].
Finally sect.~10 is devoted to discussions and future prospects.

For other approach to the 2D gravity using the collective coordinates
and supersymmetric case, we refer the reader to refs.~\cite{JEV,JRO,DMW}.

\sect{Conformal anomaly and Liouville theory}
\indent

Let us consider the $2D$ gravity coupled to a CFT with central charge
$c^M$. The partition function is given by
\EQ
Z=\int \frac{{\cal D}g{\cal D}_g X}{V(Diff)}e^{-S(X,g)-S(g)},
\EN
where $g_{ab}$ is the two-dimensional metric and $X$ represents the
matter field. We assume that the action $S(X,g)$ is invariant under
the diffeomorphism as well as the Weyl rescaling of the metric
$g \to e^\sigma g$:
\EQ
S(X,e^{\sigma}g)=S(X,g).
\EN
The action for gravity is just given by the cosmological term
\EQ
S(g)=\mu_0^2\int d^2\xi\sqrt{g},
\EN
where $\xi^a$ ${(a=1,2)}$ are the coordinates.

It is generally believed that there is no physical degrees of freedom
in the metric in two dimensions.\footnote{
The degrees of freedom in the metric in $N$ dimensions are obtained
as follows. The second-rank symmetric tensor has $N(N+1)/2$ components.
The diffeomorphism invariance subtracts $N$ and the gauge fixing
subtracts another $N$, leaving $N(N-3)/2$ degrees of freedom (Recall
the similar counting in the gauge theory). This gives $-1$ for $N=2$,
meaning no degrees of freedom in two dimensions. This is reflected
in the fact that the usual Einstein action is a total divergence and
gives just the Euler number $\frac{1}{4\pi}\int d^2 z\sqrt{g}R=-2(h-1)$,
where $h$ is the number of handles.}
We are now going to see that this is violated by the quantum effects
called conformal anomaly and this leaves nontrivial dynamics
unless certain conditions are satisfied \cite{POL1,FRI}.
Let us first choose the conformal gauge
\EQ
g_{ab}=e^{\phi_0}{\hat g}_{ab},
\EN
where ${\hat g}_{ab}$ is a reference metric conformally equivalent to
Euclidean metric $\d_{ab}$. In this gauge, it is convenient to use
a complex coordinate $z$ in place of two real $\xi^a: z=\xi^1+i\xi^2$.
We then have (for ${\hat g}_{ab}=\d_{ab}$)
$$
g_{ab}(\xi)d\xi^a d\xi^b = e^{\phi_0}d^2z, \eqno(2.5a)
$$
$$
R = e^{-\phi_0}(-4\pa_z\pa^z\phi_0).\eqno(2.5b)
$$
\setcounter{equation}{5}
Namely, $g_{ab}$ have components $g_{z{\bar z}}=g_{{\bar z}z}=
\shalf e^{\phi_0}$ on this basis.

In order to separate the volume of diffeomorphism $V(Diff)$
from the integral over the metric, we consider an infinitesimal
transformation $\d z=v^{z}(z,{\bar z})$. This induces a variation
of the metric by
\EQ
(g_{ab} +\d g_{ab}) d\xi^a d\xi^b
=g_{ab}(\xi+\d\xi) d(\xi^a+\d\xi^a) d(\xi^b+\d\xi^b).
\EN
We thus get
\bea
\d g_{z{\bar z}} &=& \pa_z (g_{z{\bar z}}v^z)
 + \pa_{\bar z} (g_{z{\bar z}}v^{\bar z}), \nonumber\\
\d g_{zz} &=& 2 g_{z{\bar z}}\pa_z v^{\bar z},~~~
\d g_{{\bar z}{\bar z}} = 2 g_{z{\bar z}}\pa_{\bar z} v^z.
\ena
Using the fact that $\frac{\pa}{\pa\phi_0}g_{zz}=\frac{\pa}{\pa v^z}
g_{zz}=0$ and $\frac{\pa}{\pa\phi_0}g_{z{\bar z}}$ is essentially
identity, we can rewrite the integral over the metric as
\bea
{\cal D}g &=& {\cal D}g_{z{\bar z}}{\cal D}g_{zz}{\cal D}
 g_{{\bar z}{\bar z}} \nonumber\\
 &=& {\cal D}_g v^z {\cal D}_g v^{\bar z} {\cal D}_g \phi_0
 \frac{\pa(g_{z{\bar z}}, g_{zz}, g_{{\bar z}{\bar z}})}
  {\pa(v^z, v^{\bar z}, \phi_0)} \nonumber\\
 &\sim& {\cal D}_g v^z {\cal D}_g v^{\bar z} {\cal D}_g \phi_0
 {\rm det}(\pa^z){\rm det}(\pa_z).
\ena
The first two integral over $v^z$ and $v^{\bar z}$
gives the volume $V(Diff)$ and the last two determinants can be
written in terms of ghosts $b_{zz}$ and $c^z$. In this way
we obtain
\EQ
Z=\int {\cal D}_g\phi_0{\cal D}_gb{\cal D}_gc{\cal D}_gX
e^{-S(X,g)-S(g)-S(g,b,c)},
\EN
where
\EQ
S(g,b,c)=\int\frac{d^2\xi}{2\pi}\sqrt{g}(b_{zz}\pa^zc^z+c.c.).
\EN
Similar to the matter action, this is conformally invariant:
\EQ
S(e^{\phi_0}{\hat g}, b,c)=S({\hat g}, b,c).
\EN
Thus the conformal field $\phi_0(\xi)$ appears to decouple from the
system. However, it does not because the volume element
${\cal D}_g X {\cal D}_gb{\cal D}_gc$ depends on $\phi_0$.

If we define an action ${\tilde S}(g)$ by
\EQ
e^{-{\tilde S}(g)} \equiv e^{-S(g)}\int{\cal D}_gb{\cal D}_gc{\cal D}_gX
e^{-S(g,b,c)-S(X,g)},
\EN
the partition function is given by
\EQ
Z=e^{-{\tilde S}({\hat g})} \int {\cal D}_g\phi_0
 e^{-S_{eff}({\hat g},\phi)},
\EN
where
\EQ
S_{eff}({\hat g},\phi_0)
={\tilde S}(e^{\phi_0}{\hat g})-{\tilde S}({\hat g}).
\EN

Our next task is to find an explicit formula for $S_{eff}({\hat g},
\phi_0)$. For this purpose, we make the stress-energy tensor by varying
the effective action $\Gamma$ with respect to the metric:
\bea
T_{zz} &=& -\frac{2\pi}{\sqrt{g}}\frac{\d\Gamma}{\d g^{zz}}
 = \frac{2\pi}{\sqrt{g}}\frac{\d W}{\d g^{zz}}, \nonumber\\
T_{z{\bar z}} &=& -\frac{2\pi}{\sqrt{g}}\frac{\d\Gamma}{\d g^{z{\bar z}}}
 = \frac{2\pi}{\sqrt{g}}\frac{\d W}{\d g^{z{\bar z}}},
\ena
where $W$ is the usual generating functional for connected diagrams:
\EQ
e^W = \int{\cal D}_gb{\cal D}_gc{\cal D}_gX
e^{-S(X,g)-S(g,b,c)-S(g)+(\chi, X)+(\b, b)+(\c, c)}.
\EN
Here $(\;,\;)$ means integral over the product. Using (2.12), we have
\EQ
W=-{\tilde S}(g)+ ({\rm terms~involving~sources}).
\EN
Since only the first term depends on $\phi_0$, we have
\EQ
T_{z{\bar z}} = \frac{2\pi}{\sqrt{g}}\frac{\d{\tilde S}}{\d\phi_0}
 g_{z{\bar z}}.
\EN
The traceless part $T_{zz}$, on the other hand, has the part which
involves $X$, $b$ and $c$ but does not depend on $\phi_0$,
and the rest $T_{zz}^{\phi}$ which depends on $\phi_0$. The first
part obeys the conservation law by itself. We thus get from the
conservation law of the stress-energy
\EQ
\pa^z T^\phi_{zz}+\pa_z\left(\frac{2\pi}{\sqrt{g}}\frac{\d{\tilde S}}
{\d\phi_0}\right) =0.
\EN
The only local quantity of rank 1 involving the metric is $\pa_z R$,
hence we find
\EQ
\pa^z T^\phi_{zz}=\pa^zT_{zz}=-\frac{\la}{24}\pa_z R,
\EN
where $\la$ is a constant to be determined. Combined with (2.19),
eq.~(2.20) yields
\EQ
\frac{\d{\tilde S}}{\d\phi_0} = \frac{\la}{48\pi}\sqrt{g}(R+\mu^2).
\EN
Integrating (2.21) using (2.5), we obtain
\EQ
S_{eff}({\hat g},\phi_0)=\frac{\la}{48\pi}\int d^2\xi \sqrt{\hat g}
\left( \shalf {\hat g}^{ab}\pa_a\phi_0\pa_b\phi_0+{\hat R}\phi_0
+\mu^2 e^{\phi_0}-\mu^2 \right),
\EN
which is the desired Liouville action.

To determine $\la$, we use the Ward identity involving (2.20).
If we vary (2.20) with a variation $\d g^{ww}$ and use the
variational formula
\EQ
\d R=(-2\pa_{\bar z}\pa_z -R)\d \phi_0 +\pa_{\bar z}\pa_{\bar z}
\d g_{zz} - \pa_z\pa_z\d g^{zz},
\EN
we obtain
\EQ
\frac{1}{2\pi}\pa^z T_{zz}T_{ww}=\frac{\la}{24}\pa_z^3\d(z-w)
 +({\rm less~ singular~ terms}),
\EN
leading to
\EQ
T_{zz}T_{ww} \sim \frac{\la}{12}\pa_z^3\frac{1}{z-w}+\cdots
 =-\frac{\la}{2}\frac{1}{(z-w)^4}+\cdots .
\EN
Thus this $\la$ must be minus of the central charge of the
matter and $b,c$ ghosts:
\EQ
\la=26-c^M.
\EN

This completes the reduction of the $2D$ gravity to the Liouville
theory coupled to a CFT. If $c^M=26$ as in the critical string,
the Liouville field is decoupled from the system and one may discard it.
However,
in non-critical case one has to incorporate the effects of the Liouville
field. This is very difficult for the following reason.

The measure for the path integral over $\phi_0$ in (2.13)
is defined by the complicated norm
\EQ
\int d^2\xi\sqrt{g}(\d\phi_0)^2
=\int d^2\xi \sqrt{\hat g} e^{\phi_0}(\d\phi_0)^2,
\EN
which depends on $\phi_0$ itself nonlinearly.
This has prevented us from proper treatment of the quantum theory
of the Liouville field and non-critical strings.
We now turn to a very interesting approach based on conformal
field theory which enables us to transform the theory into a more
tractable form.

\sect{Liouville theory as free field}
\indent

The conclusion obtained in the previous section may be summarized
as follow. The partition function in the conformal gauge is originally
given by (2.9):
\EQ
Z=\int {\cal D}_g\phi_0{\cal D}_gb{\cal D}_gc{\cal D}_gX
e^{-S(X,g)-S(g)-S(g,b,c)}.
\EN
This was rewritten as (2.13)
\EQ
Z=\int {\cal D}_{e^{\phi_0}{\hat g}}\phi_0{\cal D}_{\hat g}b
 {\cal D}_{\hat g}c{\cal D}_{\hat g}X e^{-S(X,{\hat g})-S({\hat g},b,c)
 -\frac{\la}{48\pi}S_L({\hat g},\phi_0)},
\EN
which is obtained from (2.12), (2.13) and (2.22). Here we have
defined the standard Liouville action $S_L({\hat g},\phi_{0})$ by
\EQ
S_L({\hat g},\phi_0)=\int d^2\xi \sqrt{\hat g}
\left( \shalf {\hat g}^{ab}\pa_a\phi_0\pa_b\phi_0+{\hat R}\phi_0
+\mu^2 e^{\phi_0} \right).
\EN
As it stands, the path integral over $\phi_0$ is very complicated and
it is very difficult to make sense of this theory.

In order to put this system in a more tractable form,
we make the change of variables such that the measure
is independent of $\phi_0$. This will produce a Jacobian as
\EQ
{\cal D}_g\phi_0{\cal D}_gb{\cal D}_gc{\cal D}_gX
= {\cal D}_{\hat g}\phi{\cal D}_{\hat g}b
 {\cal D}_{\hat g}c{\cal D}_{\hat g}X e^{J(\phi,{\hat g})},
\EN
where ${\cal D}_{\hat g}\phi$ is the free field measure defined
by the norm
\EQ
\int d^2\xi\sqrt{\hat g}(\d\phi)^2.
\EN
This procedure is similar to the rewriting of eq.~(3.1) as
(3.2), where the ``difference" (the Jacobian) between the
two is given by the Liouville action (3.3). In analogy to this,
we assume here that the Jacobian is given by the
exponential of a renormalizable local action similar to the Liouville
one (3.3)~\cite{DDK}:
$$
e^{J(\phi,{\hat g})} = e^{-S(\phi,{\hat g})}, \eqno(3.6a)
$$
$$
S(\phi, {\hat g})= \frac{1}{8\pi}\int d^2\xi \sqrt{\hat g}
 ({\hat g}^{ab}\pa_a\phi\pa_b\phi-2Q{\hat R}\phi +4{\mu'}^2 e^{\a\phi})
$$ $$
=\frac{1}{2\pi}\int d^2z(\pa\phi{\bar\pa}\phi-\frac{1}{2}Q
 \sqrt{\hat g}{\hat R}\phi+{\mu'}^2 \sqrt{\hat g}e^{\a\phi}),\eqno(3.6b)
$$
\setcounter{equation}{6}
where $Q,\mu'$ and $\a$ are unknown coefficients due to
quantum effects.

These parameters are determined by the consistency of the above
ansatz. First, let us choose the bare cosmological constant
$\mu_0$ so as to cancel $\mu '$. Next, to determine $Q$, notice
that the original theory depends only on $g=e^{\a \phi}{\hat g}$
and so is invariant under
\EQ
{\hat g}\to e^{\sigma}{\hat g}, \hspace{5mm} \phi\to \phi-\sigma/\a.
\EN
This means
\EQ
{\cal D}_{\hat g}\phi{\cal D}_{\hat g}b{\cal D}_{\hat g}c
{\cal D}_{\hat g}Xe^{-S(\phi,{\hat g})}
={\cal D}_{e^\sigma{\hat g}}(\phi-\sigma/\a){\cal D}_{e^\sigma{\hat g}}b
{\cal D}_{e^\sigma{\hat g}}c{\cal D}_{e^\sigma{\hat g}}X e^{-S(\phi
-\sigma /\a,e^\sigma{\hat g})}.
\EN
Since $(\phi-\sigma/\a)$ is an integration variable, we may write
this as
\EQ
{\cal D}_{e^\sigma{\hat g}}\phi{\cal D}_{e^\sigma{\hat g}}b
{\cal D}_{e^\sigma{\hat g}}c{\cal D}_{e^\sigma{\hat g}}X e^{-S(\phi,
 e^\sigma{\hat g})},
\EN
where we have used the fact that the measure for $\phi$ is
not changed under the shift of $\phi$ for (3.5). Eqs.~(3.8) and
(3.9) imply that the total conformal anomaly vanishes!
Notice that if we simply disregard the integration
over the Liouville mode $\phi_0$ in eqs.~(3.1)
and (3.2), we have the Liouville action as a conformal anomaly.
The important point here is that {\bf the inclusion of the (new)
Liouville field $\phi$ recovers the invariance}.

The stress-energy tensor for the Liouville field is given by
\EQ
T^\phi = -\shalf (\pa \phi )^2 - Q \pa^2\phi,
\EN
which tells us that it has the central charge $c^{L}=1+12Q^2$. Hence
the vanishing condition of the conformal anomaly
or total central charge reads
\EQ
c_M-26+1+12Q^2=0.
\EN
The other parameter $\a$ is determined by demanding that
$g=e^{\a \phi}\hat g$ be invariant under conformal transformation, or
$e^{\a \phi}$ be a conformal tensor of dimension (1,1).
The dimension of this operator is given by $-\frac{1}{2}\a(\a+2Q)$,
and we get
\EQ
\a_\pm =-Q \pm \sqrt{Q^2-2}=\frac{-\sqrt{25-c^M}\pm\sqrt{1-c^M}}
 {2\sqrt{3}}
\EN
We are now faced with the question which solution yields a theory
equivalent to $2D$ quantum gravity. The consistency with
the semiclassical limit $(c^\mu \to -\infty)$
\cite{ZAM} tells us that we should choose $\a_+$.

The above whole argument can be easily extended to supersymmetric
case~\cite{DHK}.

\sect{Perturbative calculation of the Jacobian}
\indent

In sect.~3, we have shown that once we assume the local form of the
Jacobian (3.6), consistency uniquely fixes its precise
form. In this section, we will describe a justification
of the local form of the Jacobian in the perturbative approach
\cite{DHO}.

The transformation from the measure defined by (2.26) to the measure
by (3.5) produces the formal Jacobian
\EQ
e^{{\tilde J}({\tilde g},\phi)}=|{\rm Det}(e^{\phi(z_1)/2+\phi(z_2)/2}
 \d(z_1-z_2))|^{1/2}.
\EN
In order to calculate this Jacobian, we have to regularize it.
For this purpose, we consider a family of metrics
\EQ
g(x)=e^{x\phi}{\hat g} \;, \hspace{1cm} 0\leq x\leq 1.
\EN
The infinitesimal contribution to the Jacobian
$\d J[{\hat g},\phi,x]$ as the metric charges from
$g(x)$ to $g(x-\d x)$ is, up to $\d x^2$,
\EQ
e^{\d{\tilde J}[{\tilde g},\phi,x]}
=|{\rm Det}(e^{\d x(\phi(z_1)/2+\phi(z_2)/2)} \d_{g(x)}(z_1-z_2))|^{1/2}.
\EN
With the help of the coordinate basis with states $|z,x>$
normalized as
\EQ
<z_1,x|z_2,x>=\d_{g(x)}(z_1-z_2)=\frac{1}{\sqrt{g(x)}}\d(z_1-z_2),
\EN
eq.~(4.3) is cast into
\EQ
\d{\tilde J}[{\hat g},\phi,x]=\shalf\int\sqrt{g(x)}\d x\phi(z)<z,x|z,x>.
\EN
This expression is ill-defined because $<z,x|z,x>\sim \d(0)$.
Let us then regularize this by using the heat kernel as
\EQ
\d{\tilde J}_\e[{\hat g},\phi,x]
 =\shalf\int\sqrt{g(x)}\d x\phi(z)<z,x|e^{-\e\Delta_{g(x)}}|z,x>,
\EN
where $\e$ is the regulator and $\Delta_{g(x)}$ is the
Laplacian for the metric $g(x)$.

The evolution operator
$G(z,z',\epsilon)=<z,x|e^{-\e\Delta_{g(x)}}|z',x>$ satisfies
\EQ
\lim_{\e \to 0}G(z,z',\e)=\d_{g(x)}(z-z'),
\EN
and is a solution of the differential equation
\EQ
\left(\frac{\pa}{\pa t}+\Delta_{g(x)}\right) G(z,z',t)=0.
\EN
The solution to this equation is well known and its short
time expansion is given as \cite{FRI}
\EQ
<z,x|e^{-\e\Delta_{g(x)}}|z,x>
=\frac{1}{4\pi\e}+\frac{1}{12\pi} R_{g(x)}+ O(\e),
\EN
giving\footnote{Note: $R_{e^{x\phi}{\hat g}}=e^{-x\phi}(R_{\hat g}
-2x\Delta_{\hat g}\phi)$.}
\EQ
\d{\tilde J}_\e[{\hat g},\phi,x]
 =\frac{1}{24\pi}\int\sqrt{\hat g}\phi[R_{\hat g}-2x \Delta_{\hat g}
 \phi]\d x +\frac{1}{8\pi\e}\int\sqrt{\hat g}e^{x\phi}\phi\d x.
\EN
The last area term may be renormalized into the cosmological constant
term in perturbation theory. Integrating (4.10) over $x$ gives the
local action assumed in (3.6).

It seems that no ambiguity appears in the above ``derivation".
However, the heat kernel regularization used
above is not unique; for instance, one may use the heat kernel
\EQ
e^{-\e(\Delta_{g(x)}+\b R_{g(x)})},
\EN
which is the most general expression that is diffeomorphism invariant.
The constant $\b$ then introduces the ambiguity in the coefficient of
$R\phi$ term which is left in (3.6).
Moreover, we should remember that the the reasoning and techniques
used above implicitly assume the validity of the usual perturbation.
The Jacobian (4.1) has short distance singularity and so
it is to be expected that the result is a local expression. Thus
it would be fair to say that we have partial support for the
ansatz (3.6) at present.

\sect{Discrete states in the bosonic Liouville theory}

\subsection{Preliminaries}
\indent

After this long preparation, we now come to the analysis of the
physical states in the $2D$ gravity coupled to $c^M\leq 1$ CFT.
We will treat this system as a free Liouville scalar
field coupled to CFT in a conformally invariant manner,
regarding the Liouville exponential interaction (the cosmological
constant term) as a marginal deformation; the effects of the
cosmological constant will be incorporated perturbatively.

We use the free field realization for the conformal matter.
The stress-energy tensor is given by
\EQ
T^X = -\shalf (\pa X)^2 - i\la^X \pa^2 X,
\EN
where scalar field $X$ has the mode expansion
\EQ
X(z)=q^X-i(p^X-\la^X)\ln z +i\sum_{n\neq 0}\frac{\a_n}{n}z^{-n},
\EN
with the commutation relation
\EQ
[\a_n,\a_m]=n\d_{n+m,0}, \;\; [q^X,p^X]=i.
\EN
It satisfies the Virasoro algebra with the central charge
$c_M=1-12(\la^X)^{2}$. This is very similar to the Liouville
theory (3.10) with $\la^X$ replaced with $\la^L=-iQ$.

The conformal invariance of the whole system may be succinctly
summarized by using the BRST charge
\EQ
Q_B = \oint \frac{dz}{2\pi i}c(z)(T^X(z)+ T^\phi(z)+\shalf T^{bc}(z)),
\EN
where $T^{bc}(z)$ is the stress-energy tensor for the ghosts.
The condition that the total central charge add to up to zero becomes
\EQ
(\la^X)^2+(\la^L)^2=-2,
\EN
which is equivalent to the nilpotency of the BRST charge.\footnote{
The total Virasoro generator $L_n=L_{n}^{X,\phi}+L_n^{b,c}$ is given
by the anticommutator $L_n=\{b_n, Q_B\}$. Hence one has
$$
[ L_n, L_m]=[ L_n,\{b_m,Q_B \}]
= -\{ Q_B,[b_m, L_n]\}+\{b_m,[ L_n,Q_B]\}
$$
$$
= (n-m)\{ Q_B,b_{m+n}\}+\{b_m,[\{b_n,Q_B\},Q_B]\}
= (n-m)L_{m+n}+({\rm terms~involving}~ Q_B^2),
$$
where we have used the relation $[b_m, L_n]=(m-n)b_{m+n}$.
This means that the nilpotency of $Q_B$ is equivalent to zero
total central charge.}
In terms of the mode operators, the BRST charge is given by~\cite{KOG}
\bea
Q_B &=& \sum_n c_{-n}L_n^{X,\phi}-\frac{1}{2}
 \sum_{n,m}(n-m):c_{-n}c_{-m}b_{n+m}:,\nonumber\\
L_n^{X,\phi} &=& \shalf\sum_n :(\a_m\a_{n-m}+\phi_m\phi_{n-m}):
 +(n+1)(\la^X\a_n+\la^L\phi_n),
\ena
where $\phi_n,c_n$ and $b_n$ are the mode operators for the Liouville
and ghost fields, and $\a_0=p^X-\la^X,\phi_0=p^L-\la^L$.
As usual, the BRST charge is decomposed with respect to the ghost
zero modes:
\EQ
Q_B=c_0L_0-b_0M+d.
\EN

The physical states are defined to be nontrivial ones satisfying
\EQ
Q_B|{\rm phys}>=0.
\EN
Any BRST-exact state ($Q_{B}\chi$) is trivial in the sense that it
trivially satisfies (5.8) and is excluded. This is what
is called the BRST cohomology. Since $L_0 = \{b_0, Q_B\},$ these
physical states satisfy
\EQ
L_0|{\rm phys}> = Q_Bb_0|{\rm phys}>.
\EN
Therefore, any physical states are BRST-exact unless they satisfy the
on-shell condition $L_0 = 0.$
It is convenient to reduce the zero eigenspace of $L_0$ by restricting
to the states annihilated by $b_0$.
In this space the physical state condition (5.8) reduces to
\EQ
L_0|{\rm phys}>= b_0| {\rm phys}> = d |{\rm phys}>=0.
\EN
Note that in this space $d^2=0$.

We now examine a few examples to reveal the general mechanism of the
origin of extra physical states.

\subsection{Examples and the general mechanism}
\indent

To see how extra states arises, let us consider all possible states
at level 1:
\EQ
\begin{array}{l}
N_{FP}=-1: b_{-1}|p,\downarrow>, \\
N_{FP}=0 : b_{-1}|p,\uparrow>, \; \a_{-1}|p,\downarrow>,\;
 \phi_{-1}|p,\downarrow>, \\
N_{FP}=1 : \a_{-1}|p,\uparrow>, \; \phi_{-1}|p,\uparrow>,\;
 c_{-1}|p,\downarrow>, \\
N_{FP}=2 : c_{-1}|p,\uparrow>,
\end{array}
\EN
where the ground state $|p,\downarrow>$ carries the momenta $p^X,
p^L$ and is annihilated by $b_0$. We apply $Q_B$ on these to find
\EQ
\begin{array}{l}
Q_B b_{-1}|p,\downarrow> = [(p^X-\la^X)\a_{-1}+(p^L-\la^L)\phi_{-1}]
 |p,\downarrow>, \\
Q_B b_{-1}|p,\uparrow> = [(p^X-\la^X)\a_{-1}+(p^L-\la^L)\phi_{-1}]
 |p,\uparrow> +2c_{-1}|p,\downarrow>, \\
Q_B \a_{-1}|p,\downarrow> = (p^X+\la^X)c_{-1} |p,\downarrow>, \\
Q_B \phi_{-1}|p,\downarrow> = (p^L+\la^L)c_{-1} |p,\downarrow>, \\
Q_B \phi_{-1}|p,\uparrow> = (p^L+\la^L)c_{-1} |p,\uparrow>, \\
Q_B \a_{-1}|p,\uparrow> = (p^X+\la^X)c_{-1} |p,\uparrow>, \\
Q_B c_{-1}|p,\downarrow> = 0, \\
Q_B c_{-1}|p,\uparrow> = 0.
\end{array}
\EN
The on-shell condition reads
\EQ
-\shalf[(p^X)^2+(p^L)^2]=1.
\EN

For general momenta, eq.~(5.12) shows that there are no nontrivial
physical states. However, for the special values of momenta
$p^X=\la^X,p^L=\la^L$, which are compatible with (5.13)
and (5.5), eq.~(5.12) becomes
\EQ
\begin{array}{ll}
Q_B b_{-1}|p,\downarrow> = 0, &
Q_B b_{-1}|p,\uparrow> = 2c_{-1}|p,\downarrow>, \\
Q_B \a_{-1}|p,\downarrow> = 2\la^X c_{-1} |p,\downarrow>, &
Q_B \phi_{-1}|p,\downarrow> = 2\la^L c_{-1} |p,\downarrow>, \\
Q_B \phi_{-1}|p,\uparrow> = 2\la^L c_{-1} |p,\uparrow>, &
Q_B \a_{-1}|p,\uparrow> = 2\la^X c_{-1} |p,\uparrow>, \\
Q_B c_{-1}|p,\downarrow> = 0, &
Q_B c_{-1}|p,\uparrow> = 0.
\end{array}
\EN
{}From these relations, we find that the following states are
nontrivial physical states:
\EQ
\begin{array}{ll}
b_{-1}|p,\downarrow> ; &
(\la^L \a_{-1} - \la^X \phi_{-1})|p,\downarrow>; \\
(\la^X \a_{-1}+\la^L \phi_{-1})|p,\downarrow>+ 2 b_{-1} |p,\uparrow>; &
(\la^L \a_{-1}-\la^X \phi_{-1})|p,\uparrow>.
\end{array}
\EN
Others are either BRST-exact or do not vanish.

There is another set of momenta at which similar miracle happens;
$p^X=-\la^X,p^L=-\la^L$. We find in this case that the nontrivial
states are given as
\EQ
\begin{array}{l}
\a_{-1}|p,\downarrow> ; \; c_{-1}|p,\downarrow>; \\
\a_{-1}|p,\uparrow>; \; c_{-1}|p,\uparrow>.
\end{array}
\EN

We can repeat similar analysis at level 2 and higher.
In this way we find there appear several nontrivial states at
the discrete values of momenta: these states appears only at
the fixed values of momenta and hence are quite different from
the usual particles.

The above examples already involve the general mechanism of how
these extra states appear. Under the action of $Q_B$, a state
$|\a,p>$ transforms into another state $|\b,p>$
with a coefficient which is a function of momenta
\EQ
Q_B|\a,p>=f(p^X,p^L)|\b,p>.
\EN
For general momenta, $f(p_X,p_L)$ does not vanish and these states
form the BRST doublet and are unphysical. If, however,
$f(p^X,p^L)$ happens to vanish, $|\a,p>$ becomes
a nontrivial physical state. Since $Q_B|\b,p>=0$ for general
momenta, we also have a physical state.
Thus the extra states appear by {\bf the ``decomposition" of
the BRST doublets into singlets}.
This also explains why such states always appear in the adjacent
values of the ghost number, as in (5.15) and (5.16), and in the general
case we discuss in the next subsection.

For example, take the first state in eq.~(5.12). Using
\EQ
p^L=i\sqrt{(p^X)^2+2}, \; \la^L=i\sqrt{(\la^X)^2+2},
\EN
which are obtained from (5.12) and (5.5), we find that
it is rewritten as
\EQ
Q_Bb_{-1}|p,\downarrow>=(p^X-\la^X)\left[\a_{-1}
 -\frac{p^X+\la^X}{p^L+\la^L}\phi_{-1}\right]|p,\downarrow>.
\EN
If we put $p^X=\la^X,p^L=\la^L$, we find two physical states
$b_{-1}|p,\downarrow>$ and $(\la^L\a_{-1}-\la^X\phi_{-1})|p,\downarrow>$
in (5.15) by this mechanism. We can show that all other states also
appear in this way. As we will discuss in sect.~10, this
is related to the vanishing of the null states and this decompostion
occurs in general at the levels where null states in the minimal model
exist; in this example, the state on the r.h.s. of eq.~(5.19) for
general momenta is a null state.

\subsection{General case}
\indent

Having got general idea how these extra states appear, the only
remaining task is to enumerate all possible cases when this happens.

For this purpose, it is convenient to rewrite the BRST
charge in the ``lightcone-like" variables defined as
\EQ
\begin{array}{l}
P^\pm (n) = \frac{1}{\sqrt 2}[(p^X +n \la^X)\pm i(p^L+n\la^L)],\\
p^\pm \equiv P^\pm (0)= \frac{1}{\sqrt 2} (p^X \pm ip^L),\\
q^\pm = \frac{1}{\sqrt 2}(q^X\pm i q^L),\;\;\; \a^\pm_n =
\frac{1}{\sqrt 2}(\a_n \pm i \phi_n).
\end{array}
\EN
We then assign the degrees to the mode operators as follows:
\EQ
{\rm deg}\left(\a^+_n,c_n \right) = +1, \;\;\;
{\rm deg}\left(\a^-_n,b_n \right) = -1
\EN
and $0$ to the ground state. All the states then carry definite degrees,
and the cohomology operator $d$ in (5.7) is decomposed into three
parts with definite degrees.
\EQ
d = d_0 +d_1+d_2.
\EN
where $d_0$ is given by
\EQ
d_0 = \sum_{n\neq 0} P^+(n)c_{-n}\a^-_n.
\EN
In our Fock space with definite degrees, each term with definite
degrees in $d^{2}=(d_0+d_1+d_2)^{2}=0$ is separately zero.
As a result, we have
\EQ
d^2_0=d^2_2=0, \;\; \left\{d_0,d_1\right\}=\left\{d_1,
d_2\right\}=0,\;\; d^2_1+\left\{d_0,d_2\right\}=0.
\EN

Our strategy for examining the cohomology problem consists of the
following two steps.
\begin{enumerate}
\item{Enumerate all possible nontrivial states satisfying
\EQ
d_0|\psi>=0.
\EN
}
\item{Examine if it is possible to extend the above obtained
states to satisfy
\EQ
d|\tilde \psi>=0
\EN
by adding higher degree terms.}
\end{enumerate}
The reason why we first search for the nontrivial states (5.25)
is that any nontrivial state satisfying (5.26) has $d_0$-nontrivial
state as its lowest degree term. Indeed, if we decompose
$|\tilde \psi>$ into terms of definite degree,
$|\tilde \psi>=|\psi_k>+|\psi_{k+1}>+\cdots$,
eq.~(5.26) gives $d_0|\psi_k>=0$ as the degree $k$ term.
Hence we must start from $d_0$-nontrivial states in order to construct
$d$-nontrivial states.

Now our first problem is quite easy. There are two possible cases
to be examined.

\underline{Case I. $P^{+}(n)\neq 0, P^{-}(n)\neq 0$ for all $n \in
 {\bf Z}$.}

In this case, if we define
\EQ
K \equiv \sum_{n\neq 0}\frac{1}{P^+(n)}\a^+_{-n}b_n,
\EN
$\{d_0, K\}\equiv{\hat N}$ becomes the number operator for the
oscillators. This means that any state satisfying (5.25) with nonzero
${\hat N}$ is trivial; the argument is similar to the on-shell
condition (5.9). Hence the
only nontrivial state is the ground state without mode oscillator:
\EQ
|p^X,p^L>,
\EN
satisfying on-shell condition $L_0=p^+p^-+{\hat N}=0$, {\em i.e.}
\EQ
- p^+p^-= -\shalf [(p^X)^2+(p^L)^2]=0.
\EN
This is what is called the ``tachyon".

\underline{Case II. $P^+(j)=P^-(k)=0$ for some integers $j,k\neq 0$.}

{}From the linearity of $P^{\pm}(n)$ in $n$, we have
\EQ
\begin{array}{l}
P^+(n) = \frac{1}{\sqrt 2}(\la^X+i\la^L)(n-j)
=\frac{1}{\sqrt{2}}t_+^X(j-n), \\
P^-(m) = \frac{1}{\sqrt 2}(\la^X-i\la^L)(m-k)
=\frac{1}{\sqrt{2}}t_-^X(k-n), \\
p^+=\frac{1}{\sqrt{2}}t^X_+ j, \;\;\;
p^-=\frac{1}{\sqrt{2}}t^X_- k,
\end{array}
\EN
where $t_{\pm}^X=-\la^X\pm\sqrt{(\la^X)^{2}+2}$ (we have chosen
$\la_L=i\sqrt{(\la^X)^2+2}$ ). The on-sell condition (5.10) then tells
us that $d_0$-nontrivial states exist only at the level ${\hat N}=
-p^+p^-=-P^+(0)P^{-}(0)=jk,$ and hence $j,k>0$ or $j,k<0$. The
question is what are the candidates for such states.

If we define
\EQ
K_j=\sum_{n\neq 0,j}\frac{1}{P^+(n)}\a^+_{-n}b_n,
\EN
${\hat N}_{0,j}=\{d_0,K_j\}$ becomes the operator for all the mode
operators but $\a^+_{-j},c_{-j}$ for $j>0$ ($\a^-_j, b_{j}$ for $j<0$).
These are the only operators that can produce nontrivial states. At
level $jk$, the nontrivial states are
\EQ
\left(\a^+_{-j}\right)^k| p^X, p^L >\;\; ,\;\;
c_{-j}\left(\a^+_{-j}\right)^{k-1}| p^X, p^L >,
\EN
for $j,k >0$ and
\EQ
\left(\a^-_j\right)^{-k}| p^X, p^L >\;\;,
b_j\left(\a^-_j\right)^{-k-1}| p^X, p^L >,
\EN
for $j,k <0$. $\bullet$

This exhausts all possible $d_0$-nontrivial states. For example, if
$P^{+}(j)=0$ but $P^{-}(n)\neq 0$, we have
\EQ
P^-(n) = \frac{1}{\sqrt 2} \left(\la^X-i\la^L\right) (n-\a),
\EN
with $\a$ being not integer. This gives $p^{+}p^{-}=P^{+}(0)P^{-}(0)
=-j\a$, which implies that the nontrivial state is possible only at
$j\a$. However, the only available oscillators have level $j$, which
cannot produce states at this level.

Having completed step 1, we now proceed to step 2.
Without detailed proof (for which we refer the reader to
refs.~\cite{BMP,IOH}), we summarize the main lemma necessary to
understand the final results.

{\bf Lemma.} If, for each ghost number $N_{FP}$, the cohomology of
$d_0$ is nontrivial for at most one fixed degree independent of
$N_{FP}$, then we can construct unique (up to $d$-exact term)
$d$-nontrivial state satisfying (5.26) from the $d_0$-nontrivial state
by adding higher degree terms. $\bullet$

Therefore we have discrete physical states whose lowest degree terms are
given in (5.32) and (5.33). It is clear that these states for $j=k=1$
correspond some of the states in (5.15) and (5.16).\footnote{Here we
only discussed the relative cohomology defined by (5.10), and the
states in sect.~5.2 contain absolute cohomology defined without the
restriction $b_0|{\rm phys}>=0$. Hence there are more states in
sect.~5.2.} Note that these discrete states have adjacent values of
the ghost number [$N_{FP}=0,1$ for (5.32) and $N_{FP}=0,-1$ for (5.33)],
as we pointed out in sect.~5.2.
These physical states exist at level $jk$ for any given integers $j$
and $k$, and there are an infinite number of them. The levels at which
they exist are precisely those where null states in the minimal models
exist (see, for instance, refs.~\cite{BPZ,KMA}). This is no accident.
We will discuss why this is so in sect.~10.

\sect{Discrete states in the super-Liouville theory}
\indent

In this section, we briefly discuss the supersymmetric extension of the
results in the previous section. The extension involves the introduction
of the additional fermionic partners $(\psi,\xi)$ and $(\b,\c)$
of $(X,\phi)$ and $(b,c)$. Using the lightcone-like variables similar
to (5.20), the BRST charge takes the form
\EQ
Q_B= c_0L-b_0M+d ~~~(~-\frac{1}{2}\c_0 F + 2\b_0K-\frac{1}{4}b_0\c^2_0~)
\EN
for the NS (R) sector, where
\bea
L_0 &=& p^+p^- +{\hat N}, \nonumber\\
d &=& d_0+d_1+d_2, \nonumber\\
d_0 &=& \sum_{n\neq 0} P^+(n)c_{-n}\a^-_n
-\frac{1}{2}\sum_rP^+(2r)\c_{-r}\psi^-_r.
\ena
The explicit forms of other operators are not necessary except
that $d^2=0$ in the Fock space with states of definite degrees.
The nilpotency of the BRST charge now gives the condition
\EQ
(\la^X)^2+(\la^L)^2=-1.
\EN

Our strategy for finding nontrivial physical states is the same
as in the bosonic case. The only difference is that we have
extra degrees of freedom coming from the supersymmetric partner.
These appear in the second term in $d_0$ (6.2).

Let us first examine the NS sector. Due to the additional
fermionic term, there are many possibilities to be
examined. Case I is the same as the bosonic Liouville and we have only
ground state $|p^X,p^L>$ with $p^+p^-=0$. Case II has the following
two possibilities:

\newpage
\underline{(i)Even $j$.}

In this case, it appears that $\a^{+}_{-j}$ and $c_{-j}$
($\a^{-}_{j}$ and $b_{j}$) for $j,k>0$ ($j,k<0$) can produce
nontrivial states, but the on-shell condition tells us that nontrivial
states are possible only at the level
\EQ
{\hat N}=-p^+p^- =-P^+(0)P^-(0)=\shalf jk.
\EN
and hence $j,k>0$ or $j,k<0$.
For odd $k$ the above mode operators cannot produce states at this
level. For even $k$, the nontrivial states are
\EQ
\left(\a^+_{-j}\right)^{k/2}| p^X, p^L >\;\; ,\;\;
c_{-j}\left(\a^+_{-j}\right)^{k/2-1}| p^X, p^L >,
\EN
for $j,k >0$ and
\EQ
\left(\a^-_j\right)^{-k/2}| p^X, p^L >\;\;, \;\;
b_j\left(\a^-_j\right)^{-k/2-1}| p^X, p^L >.
\EN
for $j,k<0$.

\underline{(ii) Odd $j$.}

Similarly the nontrivial states may be created by $\a^+_{-j},c_{-j},
\psi^{+}_{-j/2}$ and $\c_{-j/2}$ ($\a^-_{j},b_j,\psi^{-}_{j/2}$ and
$\b_{j/2}$) for $j,k <0$ $(j,k <0)$ at level $\shalf jk$. There are
many states starting from
\EQ
(\c_{-j/2})^k|p^X,p^L>,\hspace{4mm}
\psi^+_{-j/2}(\c_{-j/2})^{k-1}|p^X,p^L>,
\EN
the rest being obtained by replacing two $\c$'s by either $c_{-j}$
or $\a^{+}_{-j}$ and so on for $j,k>0$ (and similarly for $j,k <0$).
How this table ends depends on whether $k$ is even or odd, but
there is no essential difference in these two cases. $\bullet$

Just as in the bosonic case, there is no other case with nontrivial
$d_0$ cohomology.

Using the lemma in the previous section, we see that the ground state
tachyon for Case I and the states in (6.5) and (6.6) for Case II (i)
with even $j$ and odd $k$ can be extended to nontrivial elements
of $d$ cohomology. However, this lemma does not apply to the states
in case (ii) since there are many nontrivial states. We have
in this case:

{\bf Lemma$'$.}
If a $d_0$-nontrivial state transforms into another $d_0$-nontrivial
one under the action of $d$, those two states cannot give rise to
$d$-nontrivial state, since these form BRST doublets~\cite{IOH,BMP2}.
$\bullet$

Examining how the states in Case II (ii) transform into one another, we
can easily find that only the states
\bea
\psi^+_{-j/2}(\a^+_{-j})^{(k-1)/2}| p^X,p^L>, \nonumber\\
{} [(\a^+_{-j})^{(k-1)/2}\c_{-j/2}-
j(k-1)c_{-j}\psi ^+_{-j/2}(\a ^+_{-j})^{(k-3)/2}]| p^X,p^L>,
\ena
for odd $j,k >0$ and
\bea
\psi^-_{j/2}(\a^-_{j})^{-(k+1)/2}| p^X,p^L>, \nonumber\\
{} [(\a^-_{j})^{-(k+1)/2}\b_{j/2}-
\shalf b_{j}\psi^-_{j/2}(\a^-_{j})^{-(k+3)/2}]| p^X,p^L>,
\ena
for odd $j,k <0,$ are singlets and can produce nontrivial
$d$-cohomology.
(If $k$ is even, there is no singlet and hence no nontrivial state.)

To summarize, we have found that there are discrete states at level
$\shalf jk$ generated from (6.5), (6.6), (6.8) and (6.9)
for $j-k=$ even. We note that these are precisely the conditions
for the null states to exist~\cite{KMA} and that these states again
have adjacent values of the ghost number.

The R sector may be similarly examined~\cite{IOH}. After exactly the
same procedure, it is easy to show that there are nontrivial states
at level $\shalf jk$ with $j-k=$ odd for given integers $j$ and
$k$. The nontrivial states are generated form (6.5) and (6.6) for
odd $j$ and even $k$, and from (6.8) and (6.9) for even $j$ and odd $k$.
Again the above conditions are those for the existence of the null
states.

\sect{$c^M=-2$ topological gravity}
\indent

If the conformal matter coupled to the $2D$ gravity has
$c^M=-2$, it has been noted that the system is quite similar to the
topological gravity [27-30].

In our notation using free scalar fields, the stress-energy tensor
for such a system is given by
\EQ
T = -\shalf (\pa X)^2 - \frac{i}{2} \pa^2 X
 -\shalf (\pa \phi)^2 + \frac{3}{2} \pa^2\phi ,
\EN
If we use the standard bosonization for the bosonic ghost~\cite{FMS}
\EQ
\b=-i e^{-iX+\phi}\pa X \;, \;\;\;
\c= e^{iX-\phi},
\EN
eq.~(7.1) is rewritten as
\EQ
T=-2\b\pa\c - \pa\b\c .
\EN
The conformal dimensions of these ``ghosts" are dim.$(\b,\c)
=(2,-1)$ and this system has the central charge 26 which cancels
against $-26$ from the reparametrization ghosts $(b,c)$.

In the representation by $\b$ and $\c$, the Virasoro
generator $L_0$ is just the number operator for all the nonzero
mode operators. According to eq.~(5.9), this means that the nonzero mode
operators cannot produce nontrivial cohomology; only zero modes can
give rise to such states.

The mode expansion of the ghosts are given by
\EQ
\c(z)=\sum_n \c_n z^{-n+1} \;, \;\;
\b(z)=\sum_n \b_n z^{-n-2},
\EN
and the conformal vacuum satisfies
\EQ
\c_n|0>=\b_m|0>=0 \;\; {\rm for}\;\; n\geq 2,\; m\geq -1.
\EN
The usual choice of the ground state for the $\b -\c$ system is
\EQ
\c_n|{\tilde 0}>=\b_m|{\tilde 0}>=0 \;\; {\rm for}\;\; n\geq 1,\; m\geq 0,
\EN
which are related to (7.5) by\footnote{To see that $|\tilde 0>$
defined by (7.7) satisfies (7.6), one computes
$$
\c_n e^{\phi(0)}|0>=
\oint\frac{dz}{2\pi i}z^{n-2}\c(z) e^{\phi(0)}|0>
=\oint\frac{dz}{2\pi i}z^{n-2}e^{iX-\phi}(z) e^{\phi(0)}|0>
=\oint\frac{dz}{2\pi i}z^{n-1}:e^{iX(z)-\phi(z)+\phi(0)}:|0>
$$
which vanishes for $n>0$.}
\EQ
|{\tilde 0}>=e^{\phi(0)}|0>.
\EN
According to refs.~\cite{VEV,FUS}, the nontrivial cohomology is given
by the states
\EQ
\c_0^l|{\tilde 0}>, \; c_0\c_0^l|{\tilde 0}>, \;
<{\tilde 0}|\b_0^l, \; <{\tilde 0}|\b_0^l c_0.
\EN
This space has only one-dimensional extension (depends only on single
integer), in contrast to our analysis in the previous section.

In order to compare these results with ours, let us examine what momenta
$(p^X,p^L)$ these states carry. From the correspondence (7.2), we see
that $\c_0$ carries momenta $(p^X-\shalf, p^L-{3\over 2}i)=(1,i)$,
and hence the first state in (7.8) has the momenta
\EQ
\left( p^X-\shalf, p^L-\frac{3}{2}i \right)=(l, i(l-1)).
\EN
On the other hand, since $\la^X=\shalf$ and $\la^L={3\over 2}i$ ($t^X_+
=1, t^X_-=-2$), our analysis in sect.~5 indicates that there exist
extra discrete states at [see eq.~(5.30)]
\EQ
\begin{array}{l}
p^X = \frac{1}{\sqrt{2}}(p^+ +p^-)=\shalf(j-2k), \\
p^L = \frac{1}{\sqrt{2}i}(p^+ -p^-)=\frac{-i}{2}(j+2k).
\end{array}
\EN
Comparing (7.10) with (7.9), this seems to indicate $j=0$ and $2k=-2l-1$.
The latter condition contradicts to the fact that $k$ and $l$ are
integers. The resolution of this inconsistency lies probably in the fact
that the vacuum $|\tilde 0>$ is not unique; for example, if
we take $|\tilde 0>=e^{\phi(0)+(\phi(0)-iX(0))/2}|0>$,
which still satisfies (7.6), we get $j=0$ and $k=-l$. Thus
these states are at level 0 in our notation.

It is clear that our analysis allows for the wider space than that
considered in the representation in terms of $\b$ and $\c$; ours
includes states connected by the exponentials of $\phi_{1}$ and
$\phi_{2}$, which cannot be reached by simply multiplying $\b$ or $\c$.
This wider space has been known as the ``picture changed one" in
superstring theory~\cite{FMS}. If we have an additional constraint that
we should identify the picture changed states as in superstring,
then the nontrivial cohomology is exhausted by (7.8). It is not clear
to us at present whether we should impose such condition in the $c^M=-2$
theory.

\sect{Vertex operator representations for $c^{M}=1$ gravity}
\indent

The analysis in sects.~5 and 6 shows that there are an infinite
number of discrete physical states. However, their
concrete representations have not been given. Such representations are,
in fact, complicated for $c^{M}<1$ gravity and the complete
representations for all of them have not been known although some of
them have been given in terms of Schur polynomials~\cite{BMP,MUK}.
Fortunately it has been pointed out that $c^{M}=1$
gravity, the most interesting case from the physical point of view,
allows for rather simple representations by means of the vertex
operators~[14-19]. In this section, we summarize the representations and
show the BRST invariance of the states.

For $c^{M}<1$ theory, the representations of the physical states
are characterized by the screening operators with dimension
one~\cite{KMA,DF}:
\EQ
S_\pm (z)=: e^{it_\pm^M X(z)}:, \;\;
t_\pm^M=-\la^X \pm \sqrt{(\la^X)^2+2}.
\EN
In the limit $c^M\to 1 \;(\la^X\to 0)$, these constitute the $SU(2)$
current algebra with an additional generator $J^0$:
\EQ
J^{\pm}(z) = : e^{\pm i{\sqrt 2} X(z)}:, \;\;
J^0(z) = \frac{1}{{\sqrt 2}} i \pa X(z),
\EN
which satisfy the operator product expansion with the level $\kappa=1$:
\bea
J^+(z)J^-(w) &\sim& \frac{\kappa}{(z-w)^2}+\frac{2}{z-w}J^0(w), \nonumber\\
J^0(z)J^{\pm}(w) &\sim& \frac{\pm 1}{z-w} J^{\pm}(w), \nonumber\\
J^0(z)J^0(w) &\sim& \frac{\kappa/2}{(z-w)^2}.
\ena
The key of the simplicity of the construction of the discrete states
for $c^M=1$ is that these currents form a {\em closed} algebra.
Thus all the physical states in the $c^{M}=1$ matter theory belong
to representations of the $SU(2)$ current algebra, which are well
known~\cite{WZ}.

It is easy to see that $e^{i\sqrt{2}JX(z)}$
transforms as the highest weight of spin $J$ representation. By
repeatedly acting with the lowering operator $J^{-}$, we get the spin
$J$ multiplet $V_{J,m,}(m=J, J-1,\cdots,-J).$
Note that for $|m|=J$, $V_{J,m}$ are the standard tachyon operator
$e^{i\sqrt{2}mX(z)}$ at particular momenta.
The conformal dimension of $V_{J,m}$ is $J^{2}$.

These primary fields receive some gravitational dressing. We demand
that the dressed fields have conformal dimension (1,1) since then it
makes sense to integrate them over the surface. The operator
$e^{\a\phi}$ has the conformal dimension
$-{1\over 2}\a(\a-2\sqrt{2})$, (note: $\la^L=i\sqrt{2})$.
We see that there are two possibilities:
\EQ
V_{J,m}(z) e^{\sqrt{2}(1\pm J)\phi(z)}.
\EN
The $SU(2)$ quantum numbers $(J,m)$ are related to the integers $(j,k)$
introduced in sect.~5. To see this, note that the operator carries
momenta $(p^X,p^L-i\sqrt{2})=(\sqrt{2}m,-i\sqrt{2}(1\pm J)).$
On the other hand, we have $t^X_{\pm}=\pm\sqrt{2}$ and
$t^L_\pm =-i\sqrt{2}$. Combined with (5.30), we get
\EQ
J \equiv |\frac{j+k}{2}| \;,\;\; m \equiv \frac{j-k}{2}.
\EN

Similarly we have constructed all the states found in the analysis
in sect.~5. The result is summarized as follows~\cite{IOH}.

(1) For $j,k \in {\bf Z}_+$ and $N_{FP}=0$,
\EQ
\Psi^{(-)}_{J,m} (z) = (J^-_0)^{J-m} e^{i {\sqrt 2} J \phi^M(z)}
e^{{\sqrt 2}(1+J) \phi^L(z)},
\EN
where
\EQ
J^-_0\equiv \oint _{C_z} \frac{d \zeta}{2 \pi i} J^-(\zeta).
\EN

(2) For $j,k \in {\bf Z}_+$ and $N_{FP}=1$,
\EQ
{\tilde \Psi}^{(-)}_{J-1,m} (z)
= (J^-_0)^{J-m-1} \oint _{C_z} \frac{d \zeta}{2 \pi i}
\frac{K(\zeta)}{\zeta -z}e^{i{\sqrt 2}J \phi^M(z)}
e^{{\sqrt 2}(1+J) \phi^L(z)},
\EN
where\footnote{The caret on $c$ in (8.9) means that the zero mode $c_0$
is removed. The term proportional to $c_0$ in (8.8) (if we included $c_0$)
gives actually $c_0\Psi^{(-)}_{J,m}$, which has spin $J$ and should be
subtracted. This can be done alternatively by subtracting
$\pa c(z)\Psi^{(-)}_{J,m}$ (namely, without using ${\hat c}$).}
\EQ
K(z) \equiv {\hat c}(z) J^-(z).
\EN

(3) For $j,k \in {\bf Z}_-$ and $N_{FP}=0$,
\EQ
\Psi^{(+)}_{J,m} (z) = (J^-_0)^{J-m} e^{i{\sqrt 2}J \phi^M(z)}
e^{{\sqrt 2}(1-J) \phi^L(z)}.
\EN

(4) For $j,k \in {\bf Z}_-$ and $N_{FP}=-1$,
\EQ
{\tilde \Psi}^{(+)}_{J-1,m} (z)
= (J^-_0)^{J-m-1} \oint _{C_z} \frac{d \zeta}{2 \pi i}L(\zeta)
e^{i{\sqrt 2}(J-1/2) \phi^M(z)}e^{{\sqrt 2}(3/2-J) \phi^L(z)},
\EN
where
\EQ
L(z) \equiv b(z)e^{-i \phi^M(z)/\sqrt{2}}
e^{- \phi^L(z)/\sqrt{2}}.
\EN
The states representing the nontrivial cohomology classes are obtained
by acting the above operators (with $z=0$) on the physical vacuum
$|\la>\equiv |\la^X=0>\otimes |\la^L=i\sqrt{2}>\otimes c_1|0>_{bc}$.
Notice that $\Psi^{(\pm)}_{J,m}$ have spin $J$ whereas
${\tilde \Psi}^{(\pm)}_{J-1,m}$ have spin $(J-1)$.

These states may also be written in terms of the Schur polynomials
defined by
\EQ
\sum_{k\geq 0} S_k(x)z^k = \exp \left(\sum_{k\geq 1}x_k z^k \right).
\EN
They satisfy
$$
\frac{\pa}{\pa x_j}S_k(x)=S_{k-j}(x), \eqno(8.14a)
$$
$$
\sum_{m=1}^{k-j} m x_m S_{k-j-m}(x)=(k-j)S_{k-j}(x), \eqno(8.14b)
$$
$$
S_k(x+y)=\sum_{j=0}^k S_j(x)S_{k-j}(y), \eqno(8.14c)
$$
\setcounter{equation}{14}
which can be proved using (8.13). From (8.11), we have for case (4)
\EQ
{\tilde \Psi}_{J-1,J-1}^{(+)}(0)|\la>=
\oint_0\frac{d\zeta}{2\pi i}\zeta^{2-2J}\sum_{n\leq -1}b_n\zeta^{-n-2}
e^{-iX_+(\zeta)/\sqrt{2}} e^{-\phi_+(\zeta)/\sqrt{2}}|p^X_{\rm max},p^L>,
\EN
where we have denoted the creation operator terms by $X_{+}(\zeta)$
and $\phi_{+}(\zeta)$, and $p^X_{\rm max}=\sqrt{2}J, p^L=-i\sqrt{2}
(1-J)$. Noting that $iX_{+}(\zeta)+\phi_{+}(\zeta)=
\sum_{n>0}{1\over n}\a^{-}_{-n}\zeta^{n}$, we get
\bea
{\tilde \Psi}_{J-1,J-1}^{(+)}(0)|\la> &=&
\oint_0\frac{d\zeta}{2\pi i} \sum_{n\leq -1,k\geq 0}b_n\zeta^{-n-2J}
 S_k\left(-\frac{\a_{-m}^-}{m}\right) \zeta^k |p^X_{\rm max},p^L>
 \nonumber\\
&=& \sum_{n\geq 1}b_{-n} S_{2J-n-1}\left(-\frac{\a_{-m}^-}{m}\right)
 |p^X_{\rm max},p^L>.
\ena
In this way, all the states created by (8.6)-(8.12) may also be written
as
\EQ
\begin{array}{l}
(1) \; \Psi_{J,m}^{(-)}(0)|\la> = (J_0^-)^{J-m} |p^X_{\rm max}, p^L>,
 \;\;\; (J=(j+k)/\sqrt{2}), \\
(2) \; {\tilde \Psi}_{J-1,m}^{(-)}(0)|\la>
 = (J_0^-)^{J-m-1} \sum_{n=1}^{2J-1}c_{-n} S_{2J-1-n}\left(
 -\sqrt{2}\frac{\a_{-m}}{m}\right) |p^X_{\rm max}-\sqrt{2}, p^L>,\\
(3) \; \Psi_{J,m}^{(+)}(0)|\la> =(J_0^-)^{J-m} |p^X_{\rm max}, p^L>,
 \;\;\; (J=-(j+k)/\sqrt{2}), \\
(4) \; {\tilde \Psi}_{J-1,m}^{(+)}(0)|\la>
 = (J_0^-)^{J-m-1} \sum_{n=1}^{2J-1}b_{-n} S_{2J-1-n}\left(
 -\frac{\a_{-m}^-}{m}\right) |p^X_{\rm max}-\sqrt{2}, p^L>,
\end{array}
\EN
where $p^X_{\rm max}=\sqrt{2}J=|j+k|/\sqrt{2}$ and $p^L=-i{2+j+k\over
\sqrt{2}}$, in agreement with ref.~\cite{BMP}. [Note the difference
in the arguments of Schur polynomials in cases (2) and (4).]

It is instructive to check explicitly that these states are BRST
invariant. Take, for example, the state (2). Since $J_0^-$ commutes
with the BRST charge, it is enough to show this for $m=J-1$. Applying
the BRST charge on the state, we get
\bea
&& Q_B {\tilde \Psi}_{J-1,J-1}^{(-)}(0)|\la> \nonumber\\
&=& \left[\sum_{n,l>0}c_{-n}c_{-l}L_n^{X,\phi}S_{2J-1-l}
 -\shalf\sum_{n,m,l>0}(n-m):c_{-n}c_{-m}b_{n+m}:c_{-l}S_{2J-1-l}
 \right] |p^X,p^L> \nonumber\\
&=& \left[\sum_{n,l>0}c_{-n}c_{-l}L_n^{X,\phi}S_{2J-1-l}
 +\shalf\sum_{m \neq l>0}(2m-l)c_{m-l}c_{-m}S_{2J-1-l}\right]|p^X,p^L>.
\ena
Since $S_{2J-1-l}$ does not depend on $\phi_n$, we get for $n>0$
\bea
&& L_n^{X,\phi}S_{2J-1-l}(x)|p^X,p^L> \nonumber\\
&=& \left\{\frac{1}{\sqrt 2}[P^+(n)+P^-(n)]\a_n
 + \shalf\sum_{m=1}^{n-1}\a_{n-m}\a_m + \sum_{m>0}\a_{-m}\a_{m+n}\right\}
 S_{2J-1-l}(x)|p^X,p^L>,\nonumber\\
\ena
where $x_n \equiv -\sqrt{2}\frac{\a_{-n}}{n}$. Using $J={j+k\over 2},
m={j-k\over 2}=J-1$, we have $j=2J-1$ and $k=1$, and hence from (5.30)
\EQ
P^+(n)=-(n-2J+1), \;\; P^-(n)= n-1.
\EN
With the help of (8.14), the second and third terms in (8.19) may
be transformed as\footnote{Note: $\a_m=m\frac{d}{d\a_{-m}}
=-\sqrt{2}\frac{d}{dx_m}$.}
\bea
\sum_{m=1}^{n-1}\frac{d}{dx_{n-m}}\frac{d}{dx_m}S_{2J-1-l}(x)
&=& \sum_{m=1}^{n-1}S_{2J-1-l-n}(x)=(n-1)S_{2J-1-l-n}(x), \nonumber\\
\sum_{m>0} mx_m \frac{d}{dx_{m+n}}S_{2J-1-l}(x)
&=& \sum_{m>0}mx_mS_{2J-1-l-m-n}(x) \nonumber\\
&=& (2J-1-l-n)S_{2J-1-l-n}(x).
\ena
Substituting these into (8.19) yields
\bea
&& \sqrt{2}(J-1)(-\sqrt{2})\frac{d}{dx_n}S_{2J-1-l}(x)
+ (n-1)S_{2J-1-l-n}(x)+(2J-1-l-n)S_{2J-1-l-n}(x) \nonumber\\
&=& -lS_{2J-1-l-n}(x).
\ena
Putting this into (8.18), we are left with
\bea
&&Q_B {\tilde \Psi}_{J-1,J-1}^{(-)}(0)|\la> \nonumber\\
 &=& \left[-\sum_{n,l>0}l c_{-n}c_{-l}S_{2J-1-l-n}
 +\shalf\sum_{n,m>0}(m-n)c_{-n}c_{-m}S_{2J-1-m-n}\right]|p^X,p^L> \nonumber\\
 &=& -\shalf \sum_{m,n>0}(m+n)c_{-n}c_{-m}S_{2J-1-m-n}|p^X,p^L>=0.
\ena
The last equality follows from the symmetry of the sum.
It is easy to check the BRST invariance of the other states.

Finally we summarize the representation for ${\hat c}^{M}=1$
super-Liouville theory~\cite{IOH}.
These are obtained by noting that the physical states form
again representations of the $SU(2)$ current algebra generated by
\EQ
J^{\pm}(z) = :{\sqrt 2}\psi(z) e^{\pm i X(z)}:, \;\;
J^0(z) = i \pa X(z),
\EN
which satisfy (8.3) with level $\kappa =2$.

For the NS sector, they are generated by the following operators:

(1) For $j,k \in {\bf Z}_+$ and $N_{FP}=0$,
\EQ
\Psi^{(-)}_{J,m} (z) = (J^-_0)^{J-m}e^{iJ X(z)} e^{(1+J) \phi(z)}.
\EN

(2) For $j,k \in {\bf Z}_+$ and $N_{FP}=1$,
\EQ
{\tilde \Psi}^{(-)}_{J-1,m} (z)
= (J^-_0)^{J-m-1} \oint _{C_z} \frac{d \zeta}{2 \pi i}
\frac{K(\zeta)}{\zeta -z}e^{iJ X(z)} e^{(1+J) \phi(z)},
\EN
where
\EQ
K(z) \equiv [\shalf \c(z) + c(z) \psi(z)]e^{-iX(z)}.
\EN

(3) For $j,k \in {\bf Z}_-$ and $N_{FP}=0$,
\EQ
\Psi^{(+)}_{J,m} (z) = (J^-_0)^{J-m}e^{iJ X(z)} e^{(1-J) \phi(z)}.
\EN
Here $J\equiv |\frac{j+k}{2}|$ and $m \equiv \frac{j-k}{2}$ are integers.
We do not have explicit representation for case (4) at present.
It seems necessary to use complicated picture changing to construct the
states in (4). In terms of the Schur polynomials, these are written as
follows:
\EQ
\begin{array}{l}
(1) \; (J_0^-)^{J-m} |p^X_{\rm max}, p^L>, \;\;\; (J=(j+k)/\sqrt{2}),
 \nonumber\\
(2) \; (J_0^-)^{J-m-1} \left[\shalf \sum_{r>0}\c_{-r} S_{J-r-\shalf}
  \left( -\frac{\a_{-m}}{m}\right) \right. \nonumber\\
{}~~~~~~~~~~~~~~~~~~~\left.+ \sum_{n,r>0}c_{-n}\psi_{-r} S_{J-n-r-\shalf}
 \left(-\frac{\a_{-m}}{m}\right)\right]|p^X_{\rm max}-1, p^L>,\nonumber\\
(3) \; (J_0^-)^{J-m} |p^X_{\rm max}, p^L>, \;\;\; (J=-(j+k)/\sqrt{2}),
\end{array}
\EN
where $p^X_{\rm max}=J$ and $p^L=-i(1+J)$.

The BRST invariance of these states may be similarly checked. For
example, if we apply the BRST charge to the state in (2), we get,
after some algebra,
\EQ
\begin{array}{l}
Q_B {\tilde\Psi}^{(-)}_{J-1,J-1}(0)|\la> \\
= \left[\shalf \sum_{n,r>0}c_{-n}\c_{-r}(J-1)\a_n S_{J-r-\shalf}
 \left(-\frac{\a_{-m}}{m}\right)
+\frac{1}{4}\sum_{n,r>0,q}c_{-n}:\a_{-q}\a_{q+n}:\c_{-r}
 S_{J-r-\shalf}\left(-\frac{\a_{-m}}{m}\right) \right.\\
-\frac{1}{4}\sum_{n>s,r>0}\c_{-s}\c_{-r}\psi_{s-n}\a_n
 S_{J-r-\shalf}\left(-\frac{\a_{-m}}{m}\right)
-\shalf \sum_{r>n>0}\left(\frac{3}{2}n-r\right)c_{-n}\c_{n-r}
 S_{J-r-\shalf}\left(-\frac{\a_{-m}}{m}\right) \\
+\sum_{n,r,q>0}c_{-q}c_{-n}\psi_{-r}(J-1)\a_q
 S_{J-n-r-\shalf}\left(-\frac{\a_{-m}}{m}\right) \\
+\shalf\sum_{p,n,r>0,q}c_{-p}:\a_{-q}\a_{q+p}:c_{-n}\psi_{-r}
 S_{J-n-r-\shalf}\left(-\frac{\a_{-m}}{m}\right) \\
+\shalf\sum_{p,q,r>0}(q-p)c_{-p}c_{-q}\psi_{-r}
 S_{J-p-q-r-\shalf}\left(-\frac{\a_{-m}}{m}\right) \\
-\frac{1}{4}\sum_{q,n,r,s>0}(2s+q)c_{-q}c_{-n}:\psi_{q+s}\psi_{-s}:
 \psi_{-r} S_{J-n-r-\shalf}\left(-\frac{\a_{-m}}{m}\right) \\
+\shalf\sum_{n,r,s>0}\c_{-s}c_{-n}(J-1)\psi_s\psi_{-r}
 S_{J-n-r-\shalf}\left(-\frac{\a_{-m}}{m}\right) \\
+\frac{1}{2}\sum_{n,r,s>0,q}c_{-n}\c_{-s}\psi_{s-q}\a_q\psi_{-r}
 S_{J-n-r-\shalf}\left(-\frac{\a_{-m}}{m}\right) \\
-\left.\frac{1}{4}\sum_{r,s,t>0}\c_{-s}\c_{-t}\psi_{-r}
 S_{J-r-s-t-\shalf}\left(-\frac{\a_{-m}}{m}\right) \right]|p^X,p^L>.
\end{array}
\EN
We collect the terms involving $c_{-n}\c_r$ and use eqs.~(8.14) to find
\EQ
\begin{array}{l}
-\shalf\sum_{n,r>0}(J-1)c_{-n}\c_{-r}
 S_{J-n-r-\shalf}\left(-\frac{\a_{-m}}{m}\right)
+\frac{1}{4}\sum_{n,r>0,0>q>-n}c_{-n}\c_{-r}
 S_{J-n-r-\shalf}\left(-\frac{\a_{-m}}{m}\right) \\
-\frac{1}{2}\sum_{n,r,q>0}c_{-n}\c_{-r}\a_{-q}
 S_{J-n-q-r-\shalf}\left(-\frac{\a_{-m}}{m}\right)
-\shalf\sum_{n,r>0}\left(\frac{n}{2}-r\right)c_{-n}\c_{-r}
 S_{J-n-r-\shalf}\left(-\frac{\a_{-m}}{m}\right) \\
+\shalf\sum_{n,r>0}(J-1)c_{-n}\c_{-r}
 S_{J-n-r-\shalf}\left(-\frac{\a_{-m}}{m}\right)
-\shalf\sum_{s,n,r,q>0}c_{-n}\c_{-s}\psi_{s-q}\psi_{-r}
 S_{J-n-q-r-\shalf}\left(-\frac{\a_{-m}}{m}\right) \\
+\shalf\sum_{s,n,r,q>0}c_{-n}\c_{-s}\psi_{s+q}\psi_{-r}\a_{-q}
 S_{J-n-r-\shalf}\left(-\frac{\a_{-m}}{m}\right) \\
=-\shalf\sum_{n,r,s,t>0}c_{-n}\c_{-s}\psi_{-t}\psi_{-r}
 S_{J-n-r-s-t-\shalf}\left(-\frac{\a_{-m}}{m}\right)
=0,
\end{array}
\EN
where the last equality follows from the symmetry. It is easy to show
that the remaining terms also vanish. We can similarly show the BRST
invariance of other states.

States for the R sector contain the two-dimensional spinors from the
fermion zero-modes $\psi^{\pm}_{0}=\frac{1}{\sqrt{2}}(\psi_0\pm i\xi_0)
\equiv\frac{1}{2}(\sigma_{1}\pm i\sigma_{2})$. From the condition
$F \equiv 2[\b_0, Q_B]=0$, we find the spinor structure of the
state $\left( \begin{array}{c} 0 \\ 1 \end{array} \right)$ for cases
(1) and (2) and $\left( \begin{array}{c} 1 \\ 0 \end{array} \right)$
for case (3)~\cite{IOH}. So we may take the
vacuum $\left( \begin{array}{c} 0 \\ |\la> \end{array} \right)$, with
$\b_0|\la >=0$, and create the representatives of nontrivial
cohomology classes by using the same operators as in the NS sector
given in (8.25,26) for the cases (1) and (2) (but with half-odd-integers
$J$ and $m$). For case (3), we should take the vacuum $\left(
\begin{array}{c} |\la > \\ 0 \end{array} \right)$. Of course,
the mode expansions should be modified accordingly. We can similarly
check the BRST invariance of these states.

\sect{Interactions of the discrete states}
\indent

Using the vertex operator representations given in the previous section,
it is easy to examine the three-point interactions of the discrete
states for $c^{M}=1$ theory [15,16,18,19,31-35].
These can be most easily obtained from the operator product expansion
\EQ
\Psi^{(+)}_{J_1,m_1}(z) \Psi^{(+)}_{J_2,m_2}(0)
= \cdots +\frac{1}{z} C_{J_1,m_1,J_2,m_2}^{J_3,m_3}g(J_1,J_2)
 \Psi^{(+)}_{J_3,m_3}(0)+\cdots,
\EN
where $C$ are the Clebsch-Gordan coefficients and
$g(J_1,J_2)$ is an unknown function. From the dependence of the zero
modes of $X$ and $\phi$, $J_3$ and $m_3$ are determined to be
$J_3=J_1+J_2-1,\ m_3=m_1+m_2$. For these values, the coefficients
become
$$
C_{J_1,m_1,J_2,m_2}^{J_3,m_3} =
\frac{N(J_3,m_3)}{N(J_1,m_1)N(J_2,m_2)}
 \frac{J_2m_1-J_1m_2}{\sqrt{J_3(J_3+1)}}, \eqno(9.2a)
$$
$$
N(J,m) = \sqrt{\frac{(J-m)!(J+m)!}{(2J-1)!}}. \eqno(9.2b)
$$
\setcounter{equation}{2}
To determine $g(J_1, J_2)$, we may compute the operator product
for $m_1=J_1-1$ and $m_2=J_2$. In this way, one finds
\EQ
\Psi^{(+)}_{J_1,m_1}(z) \Psi^{(+)}_{J_2,m_2}(0)
= \cdots -\frac{1}{z} \frac{\sqrt{2J_3}(2J_3-1)!}
 {\sqrt{J_1J_2}(2J_1-1)!(2J_2-1)!} \frac{N(J_3,m_3)(J_2m_1-J_1m_2)}
 {N(J_1,m_1)N(J_2,m_2)} \Psi^{(+)}_{J_3,m_3}(0)+\cdots.
\EN
Hence after appropriate change of the normalization, the algebra is
given by~\cite{KP}
\EQ
{\Psi^{(+)}_{J_1,m_1}}'(z) {\Psi^{(+)}_{J_2,m_2}}'(0)
 =\cdots +\frac{1}{z}(J_2m_1-J_1m_2) {\Psi^{(+)}_{J_3,m_3}}'(0)+\cdots.
\EN
The others may be computed similarly, with the result~\cite{KP}
\bea
{\Psi^{(-)}_{J_1,m_1}}'(z) {\Psi^{(-)}_{J_2,m_2}}'(0) &=& 0, \nonumber\\
{\Psi^{(+)}_{J_1,m_1}}'(z) {\Psi^{(-)}_{J_2,m_2}}'(0) &=& 0, \;\;
 (J_1\geq J_2+1) , \nonumber\\
{\Psi^{(+)}_{J_1,m_1}}'(z) {\Psi^{(-)}_{J_3,-m_3}}'(0)
 &=& \frac{1}{z}(J_2m_1-J_1m_2) {\Psi^{(-)}_{J_2,-m_2}}'(0).
\ena
These relations are equivalent to the following three-point function:
\EQ
<0|{\Psi^{(+)}_{J_2,m_2}}'(z_1)c(z_1){\Psi^{(+)}_{J_1,m_1}}'(z_2)
c(z_2){\Psi^{(-)}_{J_3,-m_3}}'(0)c(0)|0>=(J_2m_1-J_1m_2).
\EN
The coefficient appearing in the algebra (9.4) is known to be the
structure constant of the area-preserving diffeomorphism~\cite{WIN}.

The algebra for the states with nonzero ghost number may be computed
similarly~\cite{OHS}. We have
\bea
{\tilde \Psi}^{(+)}_{J_1-1,m_1}(z){\tilde \Psi}^{(-)}_{J_3-1,-m_3}(0)
&=& \cdots +\frac{1}{z} F_{J_1-1,m_1,J_3-1,-m_3}^{J_2,-m_2}
 \Psi^{(-)}_{J_2,-m_2}(0)+\cdots , \nonumber\\
F_{J_1-1,m_1,J_3-1,-m_3}^{J_2,-m_2} &=&
C_{J_1-1,m_1,J_3-1,-m_3}^{J_2,-m_2}g(J_1,J_2),
\ena
where
\EQ
C_{J_1-1,m_1,J_3-1,m_3}^{J_2,-m_2} =
\frac{(-1)^{J_1-1-m_1}N(J_3-1,m_3)}{N(J_1-1,m_1)N(J_2,m_2)}
 [m_1J_2-m_2(J_1-1)].
\EN
To determine $g(J_1, J_2)$, we consider the special case
$m_1=-J_1+2$ and $m_2=-J_2$. After some calculation, one finds
\EQ
F=\frac{(-1)^{J_1-1-m_1}N(J_3-1,m_3)}{N(J_1-1,m_1)N(J_2,m_2)}
 \frac{(2J_3-1)!}{(2J_1-3)!(2J_2-1)!\sqrt{2J_2(J_1-1)(J_3-1)}}
[m_1J_2-m_2(J_1-1)].
\EN

Instead of continuing this line, we have computed the three-point
function
\EQ
<0|\Psi^{(+)}_{J_2,m_2}(z_1)c(z_1){\tilde \Psi}^{(+)}_{J_1-1,m_1}(z_2)
c(z_2){\tilde \Psi}^{(-)}_{J_3-1,-m_3}(0)c(0)|0>
\EN
in ref.~\cite{OHS} and found that this is given precisely by the
constant $F$ in (9.9). After changing the normalization, one finds
\EQ
<0|\Psi^{(+)}_{J_2,m_2}(z_1)c(z_1){\tilde \Psi}^{(+)}_{J_1-1,m_1}(z_2)
c(z_2){\tilde \Psi}^{(-)}_{J_3-1,-m_3}(0)c(0)|0>
= J_2m_1-(J_1-1)m_2,
\EN
again the structure constant of the area-preserving diffeomorphism.

The results (9.6) and (9.11) may be summarized by the effective
action for the three-point interactions
\bea
S_3 &=& g_0\sum_{J_1,m_1,J_2,m_2,A,B,C} f^{ABC}\left\{\shalf
 [J_2m_1-J_1m_2] g^{(-),A}_{J_1+J_2-1,-m_1-m_2} g^{(+),B}_{J_1,m_1}
 g^{(+),C}_{J_2,m_2} \right. \nonumber\\
 &-& \left.[J_2m_1-(J_1-1)m_2]{\tilde g}^{(-),A}_{J_1+J_2-2,-m_1-m_2}
 {\tilde g}^{(+),B}_{J_1-1,m_1}g^{(+),C}_{J_2,m_2}\right\}\int d\phi,
\ena
where we have introduced the open string coupling constant $g_0$
and the Chan-Paton index $A$ in the adjoint representation of
a Lie group, and associated coupling constants
$g^{(s),A}_{J,m}(\tilde g^{(s),A}_{J,m}$) with vertex operators
$\Psi^{(s),A}_{J,m}(\tilde \Psi^{(s),A}_{J,m})$ with $s=\pm$.

This can be rewritten in terms of fields defined as
\bea
\Phi(\phi,\theta,\varphi) &=& \sum_{s,A,J,m}T^Ag^{(s)A}_{J,m}M^s(J,m)
 D^J_{m,0}(\varphi,\theta,0)e^{(sJ-1)\phi}, \nonumber\\
{\tilde \Phi}^{(\pm)}(\phi,\theta,\varphi) &=& \sum_{A,J,m}T^A
 {\tilde g}^{(\pm)A}_{J-1,m}M^{\pm}(J-1,m) D^{J-1}_{m,0}(\varphi,\theta,0)
 e^{(\pm J-1)\phi},
\ena
where $M^s$ are normalization constants defined by
\EQ
M^+(J,m)=\frac{(J-1)!}{\sqrt{(2J-1)!}}N(J,m), \hspace{1cm}
M^-(J,m)=\frac{(-1)^m}{4\pi}\frac{(2J+1)\sqrt{(2J-1)!}}{(J-1)!N(J,m)}.
\EN
Note that the fields with $N_{FP}\neq 0$ have opposite statistics to
those with $N_{FP}=0$.
If we use Poisson brackets for the rotation matrix,
\bea
&&\{D^{J_1}_{m_1 0},D^{J_2}_{m_2 0}\} \nonumber\\
&=& i\frac{N(J_3,m_3)}{N(J_1,m_1) N(J_2,m_2)}
 \sqrt{\frac{(2J_1-1)!(2J_2-1)!}{(2J_3-1)!}}\frac{(J_3-1)!}
 {(J_1-1)!(J_2-1)!}(J_2m_1-J_1m_2)D^{J_3}_{m_3,0}, \nonumber\\
\ena
it is easy to show that $S_3$ can be written as\footnote{
The factors in the effective action (9.16) are obtained as follows.
Since the $\phi$ integration restricts the superscripts to $(-,+,+)$ and
produces factor three, the first term becomes ${\rm Tr}\left(\Phi^{(-)}
\left[\frac{\pa\Phi^{(+)}}{\pa\theta}\frac{\pa\Phi^{(+)}}{\pa\varphi}
-\frac{\pa\Phi^{(+)}}{\pa\varphi}\frac{\pa\Phi^{(+)}}{\pa\theta}\right]
\right)= {\rm Tr}\left(\Phi^{(-),A}\left[\frac{\pa\Phi^{(+),B}}
{\pa\theta}\frac{\pa\Phi^{(+),C}}{\pa\varphi}-\frac{\pa\Phi^{(+),B}}
{\pa\varphi}\frac{\pa\Phi^{(+),C}}{\pa\theta}\right]T^A\shalf [T^B,T^C]
\right)$, which gives the first term in (9.12). The second term is
obtained similarly.}
\EQ
S_3=ig_0\int d\phi e^{2\phi}\int_{S^2}d^2x \varepsilon^{ij}
 \mbox{Tr}\left( \frac{1}{3}\Phi\frac{\pa \Phi} {\pa x^i}
 \frac{\pa \Phi}{\pa x^j} - 2{\tilde \Phi}^{(-)}\frac{\pa{\tilde
 \Phi}^{(+)}} {\pa x^i} \frac{\pa \Phi}{\pa x^j}\right),
\EN
where $x^{i}=(\theta,\varphi)$.

This form of the effective action reminds us of the similar structure
of the nonabelian gauge theory. Following this analogy, it
is natural to look for similar "BRST-like" symmetry in the action.
For this purpose, it turns out to be convenient to write the
action for the states without ghost number in terms of the fields
\EQ
\Phi^{(\pm)}(\phi,\theta,\varphi) = \sum_{s,A,J,m}T^Ag^{(\pm)A}_{J,m}
M^\pm(J,m) D^J_{m,0}(\varphi,\theta,0)e^{(\pm J-1)\phi}
\EN
We have
\EQ
S_3=ig_0\int d\phi e^{2\phi}\int_{S^2}d^2x
 \varepsilon^{ij} \mbox{Tr}\left( \Phi^{(-)}\frac{\pa \Phi^{(+)}}
 {\pa x^i} \frac{\pa \Phi^{(+)}}{\pa x^j} - 2{\tilde \Phi}^{(-)}
 \frac{\pa{\tilde \Phi}^{(+)}} {\pa x^i} \frac{\pa \Phi^{(+)}}
 {\pa x^j}\right),
\EN
which has the symmetry under
\bea
\d\Phi^{(+)} &=& \la {\tilde \Phi}^{(+)}, ~~
\d{\tilde\Phi}^{(-)} = \la \Phi^{(-)},\nonumber\\
\d{\tilde\Phi}^{(+)} &=& 0, ~~ \d\Phi^{(-)} = 0.
\ena
Note that is a nilpotent transformation $\d^{2}=0$, similar
to the BRST transformation. This transformation is similar to that
generated by the operator $L(z)$ defined in eq.(8.7) except that this
changes spins of the states; the generator is given by
\EQ
Q=\oint\frac{dz}{2\pi i}b(z)e^{-iX(z)/\sqrt{2}-\phi(z)/\sqrt{2}},
\EN
which is clearly nilpotent. Indeed, it is easy to see
\EQ
Q \Psi^{(+)}_{J,m}={\tilde \Psi}^{(+)}_{J-\shalf,m-\shalf},\;\;
Q {\tilde \Psi}^{(-)}_{J-1,m}= \Psi^{(-)}_{J-\shalf,m-\shalf}.
\EN
The action can then be written as
\EQ
S_3=ig_0\int d \phi e^{2\phi}\int d^2x\varepsilon_{ij} \d{\rm Tr}
\left[{\tilde \Phi}^{(-)}\frac{\pa\Phi^{(+)}}{\pa x}
\frac{\Phi^{(+)}}{\pa x}\right].
\EN

What does this symmetry imply? We have no definite answer yet.
However this strongly suggests that these ``ghost degrees of freedom"
play the role of the ghosts in the usual string and cancel part of the
contribution from the $N_{FP}=0$ states.

There is some support to this conjecture. Bershadsky and
Klebanov~\cite{BEK} have recently computed the one-loop partition
function in $c^{M}=1$ gravity. The result is
\EQ
\frac{Z}{V_L}=\frac{1}{4\pi\sqrt{2}}\int d^2\tau
\frac{|\eta(q)|^2}{\tau_2^{3/2}} Z_M(\tau,{\bar\tau}),
\EN
where $V_L=|\ln\mu|/\sqrt{2}$, $\tau$ is the moduli parameter
integrated over the fundamental domain, $q=e^{2\pi i \tau}$, and
$Z_M(\tau,{\bar \tau})$ is the matter partition function given by
\bea
Z_M(\tau,{\bar\tau}) &=&
{\rm Tr} q^{L_0^{(M)}-c^M/24}{\bar q}^{{\bar L}_0^{(M)}-c^M/24} \nonumber\\
&=& \frac{1}{|\eta(q)|^2}\sum_{s,t} q^{(s\sqrt{2}/R+tR/\sqrt{2})^2/4}
 {\bar q}^{(s\sqrt{2}/R-tR/\sqrt{2})^2/4} \nonumber\\
&=& \frac{R}{\sqrt{2\tau_2}|\eta(q)|^2}\sum_{n,m} \exp\left(
 -\frac{\pi R^2|n-m\tau|^2}{2\tau_2}\right),
\ena
for a scalar field compactified on a circle of radius $R$. This
partition function contains the contribution from the primary fields
\EQ
\exp [ikX(z)+i{\bar k}{\bar X}({\bar z})],~~~
(k, {\bar k})=\left( \frac{s\sqrt{2}}{R}+\frac{tR}{\sqrt{2}},
 \frac{s\sqrt{2}}{R}-\frac{tR}{\sqrt{2}} \right).
\EN
If $|k|$ or $|\bar k|$ is $n/\sqrt{2}$, there are additional special
primary fields (8.4) whose existence is connected with the vanishing
of the null states (see the example (5.19) and the discussion
in sect.~10).

For $|k|\neq n/\sqrt{2}$, there are no null states and no special
primary fields, giving the Virasoro character $X_k=q^{k^{2}/2}/\eta(q)$.
For $|k|= n/\sqrt{2}$, the primary field $\exp (in X/\sqrt{2})$ has
a null descendant (which actually vanishes) of dimension
$(n+2)^2/4$. Thus we must subtract the latter contribution:
\EQ
\chi_{n,0}=\frac{q^{n^2/4}-q^{(n+2)^2/4}}{\eta(q)}.
\EN
However we have another primary field (of the same dimension) which
itself has a vanishing descendant. They give the character
\EQ
\chi_{n,1}=\frac{q^{(n+2)^2/4}-q^{(n+4)^2/4}}{\eta(q)}.
\EN
We have the similar sequence of the characters. If we add these
characters, all the contributions from the extra primary fields
expect those of the form (9.25) cancel out, giving
\EQ
\sum_{l=0}^{\infty}\chi_{n,l}(q)=\frac{q^{n^2/4}}{\eta(q)}.
\EN
and it gives the result (9.24). Namely, the partition function looks
as if the only primary fields are (9.25), and their Virasoro modules
do not contain vanishing descendant.

This result suggests that the ``ghost states" with spin $(J-1)$
cancel against the contributions of the discrete states with
$N_{FP}=0$ and spin $J$ except those of the ``boundary states"
$V_{J,\pm J}=\exp(\pm i JX/\sqrt{2}).$ To really check this possibility
in our approach, we have to compute the one-loop partition function with
our effective action.

It is possible to extend the above analysis to the super-Liouville
theory \cite{IOH,BMP2,KMS,OST,BS}.
There again we find the interactions of the discrete states (in the NS
sector) are governed by the same area-preserving diffeomorphism.
We refer the reader to the second reference in \cite{IOH} for the
details of computation.

\sect{Discussions}
\indent

We have shown in sect.~5 and 6 that there are extra discrete states
at the fixed values of momenta. These momenta precisely correspond to
the values at which special states with respect to the Virasoro algebra
appear. We now discuss why this is so.

For this purpose, let us use a free field realization of the Virasoro
algebra. In this representation, the Virasoro generators are expressed
in terms of the oscillators. Hence the states generated by the Virasoro
generators are rewritten by the oscillators. At level $N$, the relation
can be written as
\EQ
L^{-I}(\la)|t+\la>=\sum_J C_{IJ}(t+\la, \la)\a^{-J}|t+\la>
\EN
where $|t+\la>$ is a Fock vacuum with the momentum $p=t+\la$,
$L_{-I}(\la)$ and $\a_{-J}$ stand for all the independent
combinations of the Virasoro generators, and $C_{IJ}(t+\la,\la)$ is
a matrix depending on $t$ and $\la$. Clearly the oscillator
Fock space is much larger than the space spanned by the Virasoro
generators. This means that $C_{IJ}$ does not necessarily have
inverse. The criterion when this happens is given
by the Kac determinant formula:
$$
{\rm det}[C(t+\la,\la)]=\mbox{const.}\times \prod_{
\stackrel{j,k>0}{1\leq jk\leq N}}
(t-t_{(-j,-k)})^{P(N-jk)}, \eqno(10.2a)
$$ $$
{\rm det}[C(t+\la,-\la)]=\mbox{const.}\times \prod_{
\stackrel{j,k>0}{1\leq jk\leq N}}(t-t_{(j,k)})^{P(N-jk)}.\eqno(10.2b)
$$
\setcounter{equation}{2}
The first equation means that for the particular values of $t=t_{(-j,
-k)}\equiv \frac{1-j}{2}t_+ + \frac{1-k}{2}t_-$ $(t_{\pm}\equiv -\la \pm
\sqrt{\la^2+2})$, there are states at the level $jk$ in the oscillator
Fock space which cannot be constructed by the Virasoro generators:
\EQ
\sum_I L^{-I}|t+\la>=0,
\EN
which give physical states. At the zeros of (10.2b), there are again
physical primary states (``null states" for $c^M<1$)
at the same level $jk$. This is the origin why such discrete states
exist at these levels, where the decomposition of the BRST doublets
into singlets takes place owing to (10.3).

We have thus given a summary of the Liouville approach to the
$2D$ gravity. There are still some unsolved problems.
For example, we might say
that we now understand the origin of the discrete states fairly well.
However, the role of these states, in particular, those with ghost
number is not understood yet.

We hope that through this analysis, we may get new insight into the
$2D$ quantum gravity.

\vspace{5mm}
\noindent{\em Acknowledgement}

I would like to thank the organizers of this workshop, T. Muta,
J. Kodaira and T. Onogi for providing stimulating atmosphere. The work
described in sects.~6 and 8 was done in collaboration with Katsumi Itoh.
I am also grateful for useful discussions with K. Fujikawa, N. Sakai
and Y. Tanii.

\end{document}